\documentclass[twocolumn,showpacs,preprintnumbers,amsmath,amssymb]{revtex4}
\usepackage{graphicx}
\usepackage{subfigure}
\usepackage{epsfig}
\usepackage{epstopdf}
\usepackage{dcolumn}
\usepackage{bm}
\usepackage{amsmath}
\usepackage{color}
\newcommand {\apgt} {\ {\raise-.5ex\hbox{$\buildrel>\over\sim$}}\ }

\begin{document}
\title{A family of braneworld wormholes }

\author{Gabino Estevez Delgado}
\email{gestevez.@gmail.com} \affiliation{Facultad de Qu\'imico
Farmacobiolog\'ia de la Universidad Michoacana de San Nicol\'as de
Hidalgo, Tzintzuntzan No. 173, Col. Matamoros, C.P. 58240, Morelia
Michoac\'an, M\'exico.}
\author{Joaquin Estevez Delgado} \affiliation{Facultad de Ciencias
Fisico Matematicas de la Universidad Michoacana de San Nicol\'as de
Hidalgo, Ciudad Universitaria, Morelia Michoac\'an, M\'exico.}
\email{joaquin@fismat.umich.mx}
\author{ Refugio Rigel Mora-Luna}
\email{rigel@fis.unam.mx} \affiliation{Facultad de Ingenier\'ia
Qu\'imica de la Universidad Michoacana de San Nicol\'as de Hidalgo,
Edificion M, Ciudad Universitaria,   CP 58030,  Morelia Michoac\'an,
M\'exico.}

\begin{abstract}
A wormhole solution is presented on a braneworld scenery, the
material source is given by a perfect fluid on brane that does not
violate the energy conditions. we propose an ansatz for the state
equation resulting in a family of solutions that satisfy necessary
conditions over the free parameters of the solution. The effect of
projecting the 5D Weyl tensor components on  the brane, give the
conditions to have a wormhole satisfying the NEC, besides
simultaneously guaranteeing the absence of horizons. Furthermore,
our model predicts an astronomical bound for the wormhole throat.

\end{abstract}


\maketitle

\section{Introduction}

Space-time's represented of wormholes are associated with matter
fields that violate the null energy condition ($c^2\rho+P<0$)
\cite{Thorne,Morris2,Hochberg1,Idaa, Visser1} or as well as the
violation of the  null energy condition  by the  efective
stress--energy tensor \cite{Letelier,Anchordoqui,Hochberg2}. This
due to the existence of a minimal area region known like throat that
connect two spacetime regions, on static case the throat is a stable
minimum close surface embedded in the orthogonal hypersurface to the
temporal Killing vector field connecting two space time regions.
Wormholes in the beginning are the result of a theorical approach
that has attracted the attention of researches given their
attractive properties of connecting two regions of space-time and
its possible use for time travel \cite{Thorne}.

Although a numerical analysis suggest that quantum effects may be
sufficient to keep the throat of a wormhole \cite{Popov}. The
possible existence is sustained due to the observational results of
the accelerating universe \cite{Reiss,Perlmutter}, which  implies
the existence of dark energy, which violates the strong energy
condition $(c^2\rho+3P)<0$, an example of this is the called ghost
energy that have the state equation  $P=-wc^2 \rho$, $w>1$
\cite{Caldwell}, and for this state equation also we have the
violation of the null energy condition $(c^2\rho+P)<0$. Diversity of
wormholes have been presented, in some cases based on the standard
gravitational theory of Einstein and assuming a fluid, some people
have built various asymptotically flat models, or by the cutting and
paste technique having a layer associated with the wormhole
\cite{Lemos,Kuhfittig}, with matter sources and some scalar field
\cite{Barceló} or simply with scalar fields (ghost fields)
\cite{Ellis,Urazalina}. In other cases it is considered the
construction of wormholes in different gravitational theories to
Einstein gravitational theory, like $f(R)$ theory \cite{Oliveira} or
the Finsler geometry \cite{Bao,Rahaman}

One of the themes that has been developed in different researches,
is the construction of wormholes with matter sources  that does not
violate the null energy condition, but as we know the effective
stress-energy tensor in fact violate the null energy condition
\cite{Visser1}. In the $f(R)$ theory,  have been built wormholes in
which the fluid considered, anisotropic, does not violate the energy
conditions WEC and NEC, this is possible due the election of $f(R)$
and the state equation \cite{Oliveira} or  the Einstein--Cartan
theory \cite{Bronnikov} between others.

An alternative option to formulate models made by the standard GR
are the braneworlds  \cite{Maartens1}, which emerge as an
alternative geometric solution to the problem of mass hierarchy
\cite{RS1}. The braneworlds can be defined as 3+1 hypersurfaces
called branes where all the matter fields are confined except
gravity, which is free to propagate in the 5D bulk. The physics our
universe can be recovered from the braneworld, making a dimensional
reduction of the effective $5D$ theory to the $4D$, as shown in
\cite{maeda}. the $4D$  gravitational field equations on the brane
contain corrections to the GR, and therefore worth exploring new
wormhole solutions \cite{KA,lobo} and study the phenomenology that
the braneworlds bring to the problem of exotic matter.  The
gravitational field equations on the brane are recovered from the 5D
warped geometry theory, which contributes to the 4D theory with an
effective stress-energy tensor, in particular for a perfect fluid we
can recover the conventional GR, plus corrections given in terms of
$P$ and $\rho$, that contain  the 5D Weyl tensor projections, which
are responsible for providing new physics phenomenology to the
wormhole. In this case we have more freedom to play with the
effective stress-energy tensor, so we can find an alternate route on
the braneworlds to solve the problem of exotic matter, taking care
do not violate the null energy condition, these new ideas have been
seriously mentioned in other works
\cite{KA,lobo,Snath,Chakraborty,KC,Cmolina1,Cmolina2,MLcamera}.
Braneworld models allows the construction of wormholes in which the
stress-energy tensor, confined to the brane, does not violate
conditions of energy. The Weyl tensor projected components on the
brane cause the violation of the null energy condition, due to the
effective radial pressure and density and also induce anisotropic
pressures \cite{lobo}. However, it is possible to relax this
situation by a careful choice of matter on the brane. In this paper,
we will build a WH solution considering a perfect fluid on the
brane, which does not violate the null energy conditions, but
considering the corrections to the GR equations from the projections
of 5D Weyl tensor, we obtain an effective anisotropic fluid that
does not satisfy the null energy condition, in addition, in order to
build our solution, we propose an ansatz for the state equation in
function of the radial coordinate and parameters of the brane and
bulk.

\section{The Equations on the brane}
The components of the Einstein equations in five dimensions are
given by \cite{Maartens1}
\begin{equation}
^{(5)}G_{AB}=-\Lambda_5 \,^{(5)}g_{AB}+k_5^2\, ^{(5)}T_{AB} \,.
\label{E1}
\end{equation}
in our case we will not consider matter fields on the bulk
$^{(5)}T_{AB}=0$, the field equations projected on brane are given
by
\begin{equation}
^{(4)}G_{\mu\nu}=-\Lambda g_{\mu\nu}+k\, T_{\mu\nu}+
\frac{6k}{\lambda}S_{\mu\nu} +\frac {6}{k\lambda}W_{\mu\nu} \,,
\label{E2}
\end{equation}
where
\begin{equation}
k=\frac{\lambda k_5^2}{6} \,, \qquad
\Lambda=\frac{1}{2}(\Lambda_5+k\lambda) \,,
\end{equation}
where $k$ and $k_5^2$ are the gravitational coupling constants,
$\Lambda$ and $\Lambda_5$ the cosmology constants on brane and the
bulk, respectively; $\lambda$ it is the tension of the brane.
$T_{\mu\nu}$ is the stress-energy tensor confined on the brane. In
order to have a realistic braneworld scenary in concordance with
RSII, we have fixed $\Lambda_5=-k\lambda$. While that
\begin{equation}
S_{\mu\nu}=\frac{1}{12}T T_{\mu\nu}-\frac{1}{4}
T_{\mu\alpha}T^{\alpha}{}_{\nu}+
\frac{1}{8}\,g_{\mu\nu}\left[T_{\alpha\beta}T^{\alpha\beta}-\frac{1}{3}T^2
\right] \,, \label{E3}
\end{equation}
are the first correction to Einstein equations of general
relativity. Meanwhile the terms
$\frac{6}{k\lambda}W_{\mu\nu}=-\delta_\mu^A\,\delta_\nu^C\;^{(5)}C_{ABCD}\,n^Bn^D$,
are the $5D$ Weyl tensor projections that correspond to corrections
of second order on the braneworld. We assume a stress-energy tensor
of a perfect fluid
\begin{equation}
 T_{\mu\nu}= ({\it P} + c^2\rho ) u_{\mu}u_{\nu}+ {\it P} g_{\mu\nu},
\end{equation}
where the Weyl tensor projections are given by
\begin{equation}
 W_{\mu \nu}=\frac{1}{3}(4\,{\cal U}-{\cal P}) u_{\mu}u_{\nu}+{\cal P}X_{\mu}X_{\nu}+\frac{1}{3}( {\cal U}-{\cal P})
 g_{\mu \nu}.
\end{equation}
For a static and spherically symmetric space-time we will consider
the metric in Schwarzschild  coordinates
\begin{eqnarray}
ds^2\!=\!-y(r)^2\,dt^2+\frac{dr^2}{1- b(r)/r}
   +\!r^2 (d\theta ^2+\sin ^2{\theta}\,d\phi ^2),
\label{elementodelinea}
\end{eqnarray}
with the range of coordinates  $(t, \theta,\phi)$ ass the usual
$r\geq r_0$, where $r_0$ correspond to the throat and
\[
 u_{\mu}={\delta_{\mu}}^{t} y \left( r \right);
\qquad
  X_{\mu}={\frac {{\delta_{\mu}}^{r}}{ \sqrt{B \left( r \right) }}},
\]
additionally must satisfy  $b(r_0)=r_0$, $(b-b'r)/b^2>0$, or
equivalently $b'(r_0)<1$, this implies that the effective
stress-energy tensor violates the null energy condition
\cite{Thorne}, for values of  $r>r_0$, $1-b/r\geq 0$. Furthermore,
the absence of horizons is satisfied if $y(r_0)>0$.

The field equations in this coordinates for a perfect fluid have the
form:
\begin{eqnarray}\label{eq:U}
\frac {b'}{r^2}&=& kc^2\rho +\frac{k}{2\lambda}{c}^{4}  \rho ^{2}
+\frac {6 }{k\lambda}{\cal U}, \\\nonumber -\frac {b
}{{r}^{3}}&=& \left( 1-{\frac {b }{r}} \right) \frac{2y'}{ry }+ k{\it P} +{\frac {kc^2\rho ( 2\,{\it P} +c^2\rho  ) }{2\lambda}}\\
&+&\frac {4{\cal P} +2{\cal U} }{k\lambda}, \label{eq:br}
\\ \label{eq:PP} \nonumber k{\it P} &=&{\frac {( r-b)( ry''+y') }{y
{r}^{2}}}- {\frac{( y'r+y)( b'r-b)
}{2{r}^{3}y}}\\
&-&{\frac {kc^2\rho(2{\it P} +c^2\rho)}{2\lambda}} -{\frac {2({\cal
U}-{\cal P}) }{k\lambda}}, \\\label{eq:rhoprime}
 - \rho'&=&\frac {
4( 2\,{\cal U} +{\cal P}  ) y'}{{k}^{2}c^2 \left( {\it P} +c^2\rho
\right)y } +\frac {\left(4{\cal P}'+2{\cal U}'\right)r+12{\cal P}
}{{k}^{2}c^2 \left( {\it P} +c^2\rho \right)r},\\\label{eq:P}
 {\it P}'&=&-{\frac { \left( {\it P} +c^2\rho  \right) y'}{y }},
 \end{eqnarray}
where $'$ denotes the derivative with respect to the coordinate $r$.
Below in the next section we will build the solution for the
wormhole.
\section{The Solution}\label{The solution}
\subsection{Integrating the equations }
\begin{figure}[htb]
\begin{center}
\includegraphics[width=7cm]{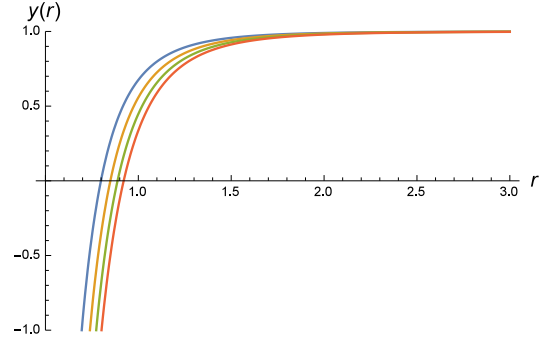}
\end{center}\vskip -5mm
\caption{The shape for $y(r)$ is traced in natural units,
considering $r_0=1$, $\mu=1$, $\lambda=0.36$, the blue, orange,
green and red lines are plotted with $\beta=3,4,5,6$ respectively.}
\label{fig:gtt}
\end{figure}

 Given the complexity to found a general analytic solution to the equations
 (\ref{eq:U}), (\ref{eq:br}), (\ref{eq:PP}) and (\ref{eq:P}),
 we have chosen to propose an ansatz for the equation of state as
 follow
\begin{equation}\label{eq:stateeq}
 {\it P} =\frac {\mu{c}^{4} \rho^{2}}{2\lambda} \left( 1-{\frac {\mu\,c^2\rho }{\lambda}} \right)
 ^{-1},
 \end{equation}
 likewise
 \begin{equation}
 \rho ={\frac {\beta\,{{\it
 {r_0}}}^{3}}{kc^2{r}^{5}}},
 \end{equation}
 where $r_0$ is the radius of the throat and $\mu$ is an arbitrary positive constant. To build the
  solution we proceed as follows: We replace (\ref{eq:stateeq}) in the conservation equation for the  perfect fluid
(\ref{eq:P}) to obtain the following equation
\begin{equation}
 {\frac {y'}{y }}={\frac {c^2\mu\,\rho'}{\mu\,c^2\rho -\lambda}},
\end{equation}
Integrating we obtain the following expression for $y(r)$
\begin{equation}\label{eq:ysol}
 y =C_{{1}} \left( 1-{\frac {\mu\,c^2\rho }{\lambda}} \right),
\end{equation}
where $C_{{1}}$ is the arbitrary constant of integration, which we
will fix as $1$ in order  to impose an asymptotically flat solution
(see Figure \ref{fig:gtt}). Replacing ${\it \rho}$ in the equation
(\ref{eq:U}) we obtain an expression for ${\cal U}(r)$
\begin{equation}\label{eq:USol}
{\cal U}={\frac {k\lambda b'}{6{r}^{2}}}-{\frac {\beta\,{{\it
{r_0}}}^{3} \left( 2\,k\lambda{r}^{5}+\beta\,{{\it {r_0}}}^{3}
\right) }{12r^{10}}} \,.
\end{equation}
Replacing (\ref{eq:stateeq}), (\ref{eq:ysol}) and (\ref{eq:USol}) in
the equation (\ref{eq:PP}) we obtain ${\cal P}$
\begin{align}
\nonumber
 {\cal P} =&
\frac{\beta\,{{\it {r_0}}}^{3} \left(
2\,{\lambda}^{2}{k}^{2}{r}^{10}-\beta\,{{\it {r_0}}}^{3} \left(
5\,\mu+2 \right) \lambda\,k{r}^{5}-\mu\,{\beta}^{2}{{\it {r_0}}}^{6}
\right) }{24{r}^{10} \left( k\lambda\,{r}^{5}-\mu\,\beta\,{{\it
{r_0}}}^{3} \right) }  \\
&+\frac{5 \left( r-b  \right) \lambda\,\mu\,\beta\,{{\it
{r_0}}}^{3}k}{2{r}^{3} \left( k\lambda\,{r}^{5}-\mu\,\beta\,{{\it
{r_0}}}^{3} \right) } -{\frac{k\lambda\, \left( b'r+3\,b  \right)
}{12{r}^{3}}} \label{eq:PSol} \,.
\end{align}
Replacing (\ref{eq:stateeq}), (\ref{eq:ysol}) and (\ref{eq:PSol})
(\ref{eq:USol}) in the equation (\ref{eq:br}) we obtain the next
differential equation:
\begin{align}\label{eq:bprime}\nonumber
 & \left( 2k\lambda{r}^{5}+3\,\mu\,\beta\,{{\it {r_0}}}^{3} \right)b'
  -45{\frac {\beta\,\mu\,{{\it {r_0}}}^{3}b }{r}}+\beta{{\it {r_0}}}^{3}(
  40\mu-k\lambda\,{r}^{2})\\
&+\frac{{\beta}^{2}{{\it {r_0}}}^{6} \left( 5\,\mu+2 \right)
}{2{r}^{3}}+{\frac {{\beta}^{3}{{\it
{r_0}}}^{9}\mu}{2k\lambda\,{r}^{8}}}=0 \,.
\end{align}
Integrating we obtain the expression for $b(r)$
\begin{align}\label{eq:b}\nonumber
 b &=-\frac {{r}^{15}{ C_2}}{ \left(\vphantom{} 2k\lambda{r}^{5}+3\mu\,\beta\,{{\it {r_0}}}^{3} \right) ^{3}}
 \\\nonumber &+\frac {1}{ \left( 2\,k\lambda\,{r}^{5}+3\,\mu\,\beta{{\it {r_0}}}^{3}
  \right) ^{3}}
\!\left[\vphantom{\frac {180\,{\beta}^{3}{{\it
{r_0}}}^{9}{\mu}^{3}r}{7}}\!\!-2\,{r}^{11}{\lambda}^{2}\beta\,{{\it
{r_0}}}^{3}{k}^{2} \!\left( k\lambda\,{r}^{2}-20\,\mu \right)\right.
\\\nonumber &-\frac{2\,{r}^{6}\lambda\,{\beta}^{2}{{\it {r_0}}}^{6}k\left(
3\,k\lambda\, \left( \mu-2 \right) {r}^{2}-560\,{\mu}^{2}
\right)}{21} +{\frac {9\,{\beta}^{5}{{\it
{r_0}}}^{15}{\mu}^{3}}{44\,k{r}^{7}\lambda}}\\\nonumber
 &+{\frac {7\,{\beta}^{3}k\lambda\,\mu\,{{\it {r_0}}}^{9} \left( 3\,\mu+2 \right)
 {r}^{3}}{12}}+\frac {15\,{\beta}^{4}{\mu}^{2}{{\it
{r_0}}}^{12} \left( 3\,\mu+2 \right) }{34\,{r}^{2}}\\
&+\left. \frac {180\,{\beta}^{3}{{\it {r_0}}}^{9}{\mu}^{3}r}{7}
\right] \,.
\end{align}
In this case the integration constant was fixed imposing the
condition  $b(r_0)=r_0$, for obtain
 \begin{align}\nonumber \label{eq:c2}
 {\it C_2}&=-2{k}^{3}{\lambda}^{3} \left( \beta+4 \right) {\it {r_0}}-{\frac {2\beta\,{k}^{2}{\lambda}^{2} \left(  \left( \mu-2 \right) \beta-14\,\mu \right) }{7{\it
 {r_0}}}}\\\nonumber
&+\frac {{\beta}^{2}k\lambda\,\mu\, \left( 7\, \left( 3\,\mu+2
\right) \beta-8\,\mu \right) }{12{{\it {r_0}}}^{3}}+\frac
{9\,{\beta}^{5}{\mu}^{3}}{44\,k\lambda\,{{\it {r_0}}}^{7}}\\
&+{\frac {3\,{\beta}^{3}{\mu}^{2} \left( 35\, \left( 3\,\mu+2
\right) \beta-102\,\mu \right) }{238\,{{\it {r_0}}}^{5}}}.
\end{align}
With this, we have managed to integrate the system, the expressions
for ${\cal U}$ and ${\cal P}$ will remain indicated in terms of
 $b$ and $b^{'}$ because they are a overlong expressions.
\subsection{Analysis of the solution}
\begin{figure}[htb]
\begin{center}
\includegraphics[width=7cm]{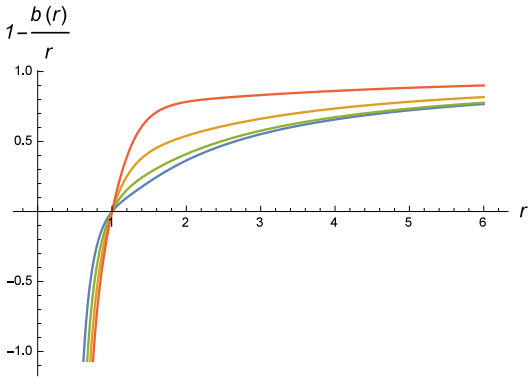}
\end{center}\vskip -5mm
\caption{The shape of $g^{rr}$ is traced in natural units,
considering $r_0=1$, $\mu=1$, $\lambda=0.36$, the blue, green,
orange and red lines are plotted with $\beta=3,4,5,6$ respectively.}
\label{fig:grr}
\end{figure}

In order to know the asymptotic behavior of our solution, we will
expand $g^{rr}$ around of $r=\infty$ as follow
\begin{align}\label{eq:expb}
1-\frac {b }{r}&=1+{\frac {{\it
C_2}}{8\,{k}^{3}{\lambda}^{3}r}}+\,\frac {\beta\,{{\it
{r_0}}}^{3}}{4\,{r}^{3}}-\,\frac {5\,\mu\,\beta\,{{\it
{r_0}}}^{3}}{k\lambda\,{r}^{5}}-O \left( {r}^{-6} \right).
\end{align}
Together with the previous expansion the expression (\ref{eq:ysol})
show that the obtained solution is asymptotically flat when
$r\rightarrow\infty$ (see Figure \ref{fig:grr}).
Near of or in the throat $b=r$ we require that our wormhole is
connected with an asymptotically flat surface, this should be
satisfied for the flare-out condition, which implies that $b^{'}<1$,
clearing $b^{'}$ from (\ref{eq:bprime}) we obtain
\begin{equation} b'=\frac { \left[2{k}^{2}{\lambda}^{2}{{\it {r_0}}}^{4}-k\lambda \left( 5\mu\beta+2\beta-10\mu \right) {{\it {r_0}}}^{2}
\mbox{}-{\beta}^{2}\mu \right] \beta}{2{{\it {r_0}}}^{2}\lambda\,k
\left( 2{{\it {r_0}}}^{2}k\lambda+3\mu\beta \right) }.
\end{equation}
Applying the flare--out condition to previous expression, and after
some algebra we obtain the following inequality
\begin{align}\nonumber
 &-\beta\, \left( {{\it {r_0}}}^{2}k\lambda\, \left( 5\,\beta-4 \right) +{\beta}^{2} \right) \mu\\
 &-2\,k\lambda\,{{\it {r_0}}}^{2} \left({\beta}^{2}-{{\it {r_0}}}^{2}k\lambda\, \left( \beta-2 \right)  \right)
<0.
\end{align}
The above condition along with the condition of nonexistence of
horizons, reveals possible bounds on the parameters in our solution.
For our convenience we will introduce the following redefinition
$x={{\it{r_0}}}^{2}k\lambda$. From the density $\rho$ in
(\ref{eq:stateeq}), we conclude that $\beta>0$, together with the
condition of absence of horizons near the throat $y(r_0)>0$ we
obtain the following restriction
\begin{equation}\label{eq:uneq1}
 \frac{\beta \, \mu}{x}<1.
\end{equation}
Under the previous redefinition we seek to satisfy conveniently the
flare—-out condition, also taking into account the absence of
horizons as shown below
\begin{align}\label{eq:uneq2}
\frac{ 2\,x \left( \beta-2 \right) -2\,{\beta}^{2}}{x \left(
5\,\beta-4 \right) +{\beta}^{2}}<\frac {\beta\,\mu\, }{x}\,.
 \end{align}
In order to use a realistic estimates for the parameter $\lambda>0$
we must demand that $\mu>0$, to do this we must also impose the
following restriction on the left side of  (\ref{eq:uneq2})
\begin{align}\label{eq:uneq3}
\frac{ 2\,x \left( \beta-2 \right) -2\,{\beta}^{2}}{x \left(
5\,\beta-4 \right) +{\beta}^{2}}>0\,.
 \end{align}
To satisfy inequality (\ref{eq:uneq3}) is necessary impose
$x>\frac{\beta ^2}{\beta -2}$, from this restriction is evident that
$\beta>2$ and considering previous restrictions, we can establish
directly from (\ref{eq:uneq1}) and (\ref{eq:uneq2}) the valid range
for the parameter $\mu$ as follow
\begin{equation}
x \left(\frac{1}{\beta }-\frac{3 (\beta +x)}{\beta ^2+(5 \beta -4)
x}\right)<\mu <\frac{x}{\beta }.
\end{equation}
It is noted that, the absence of additional zeros in the function
$g^{rr}$, should be done by a fine-tuning on the constant $\mu$,
because an extra condition for the constant $\mu$ taking into
account the regularity of $g_{rr}$ is difficult to obtain. Upon
returning to the original variable, we can set a lower bound for the
size of the throat of wormhole in terms of the brane tension and the
parameter $\beta$, as shown below
\begin{equation}\label{eq:rmin}
r_0>\frac{\beta }{\sqrt{(\beta -2) k \lambda }}\,.
\end{equation}
Although our model can be considered a toy model, we can estimate in
a first approximation the smaller radius $r_0$ permitted by our
model, considering the lower bound for
$$
\lambda \apgt 84.818\,\mathrm{MeV}^{4},
$$
reported in \cite{Aspeitia},  which meets the minimum requirements
for a real stellar setting for white dwarfs; using the condition
(\ref{eq:rmin}) we found  the minimum value allowed for $r_0\sim
98.139 \text{pc}$, it is worth noting that the minimum value $r_{0}$
can decrease if an upper bound is established for $\lambda$.

It should also be noted that the condition of absence of horizons
around the throat, along with the requirement $\lambda>0$, restricts
the radial pressure of the perfect fluid to be define positive for
$r\geq r_{0}$ as shown in the following expression
\begin{align}
 {\it P}(r)=\frac{\mu ~\beta ^{2}~{\it {r_0}} ^{6}}{2k ^{2}~r ^{10}~\left(\lambda -\frac{\mu ~\beta ~{\it {r_0}} ^{3}}{k ~r ^{5}}
 \right)}.
\end{align}
Contributions made by the  $5D$  Weyl tensor projections to the
brane decay rapidly to zero, to ensure that the anisotropic fluid
generated by the wormhole decreases rapidly away from the throat as
$r\rightarrow \infty$; below is shown the asymptotic form for ${\cal
U}$ and ${\cal P}$
\begin{equation}
{\cal U}=-{\frac {\beta{{\it {r_0}}}^{3}k\lambda}{12{r}^{5}}}-{\frac
{10\beta\mu{{\it {r_0}}}^{3}}{3{r}^{7}}} -{\frac {15\beta{{\it
{r_0}}}^{3}{\it C_2}\mu}{32{k}^{3}{\lambda}^{3}{r}^{8}}} +O \left(
{r}^{-10} \right),
\end{equation}
\begin{equation}  {\cal P}={\frac {{\it
C_2}}{32{k}^{2}{\lambda}^{2}{r}^{3}}}+{\frac {5\,\beta\,{{\it
{r_0}}}^{3}k\lambda}{48\,{r}^{5}}}
 +{\frac {35\beta\mu{{\it {r_0}}}^{3}}{12\,{r}^{7}}}+O \left( {r}^{-8}
 \right),
\end{equation}
where the constant ${\it C_2}$ is defined in (\ref{eq:c2}).
Comparing with other works in the literature, where wormhole
solutions are explore, exploiting the fact of having $R\neq 0$ using
only dust, as in the case of \cite{lobo}, our exact solution is more
general. The regularity of spacetime required as a condition of the
wormhole is satisfied as can be seen in the form of scalar curvature
\begin{align}\nonumber
 &R=-\frac{\beta\,{{\it {r_0}}}^{3}}{2\,k^2\lambda {r}^{15}y(r)
}\times\\
&\times\left(\frac{{\beta}^{2}\mu\,{{\it {r_0}}}^{6}}{\lambda}
-2\,{k}^{2}{\lambda}{r}^{10}+5\,\beta\,k\,\mu\,{r}^{5}{{\it
{r_0}}}^{3}+2\,\beta\,k\,{r}^{5}{{\it {r_0}}}^{3} \right),
\end{align}
this is regular for  $r\geq r_0$ as $y(r)>0$, meanwhile for
${r\rightarrow \infty}$ $R=0$. The Kretschmann scalar $K=R^{\alpha
\beta\mu \nu}R_{\alpha \beta\mu \nu}$ also is regular and tends to
zero when $r$ tends to infinity.

Finally, we will analyze the expression related to the energy
condition given by the following expression
\begin{equation}\label{eq:denenerg}
c^2\rho  +{\it P}  =\frac{\beta\,{{\it {r_0}}}^{3}}{k\,{r}^{5}}
\left[ 1+ \left( 1-{\frac {\beta\,\mu\,{{\it
{r_0}}}^{3}}{k\lambda\,{r}^{5}}} \right) ^{-1} \right].
\end{equation}

Analyzing the expression (\ref{eq:denenerg}) according to
(\ref{eq:rmin}), we find that, in order to have  the energy density
defined positive around the throat $r=r_0$, it requires that $\mu
<\frac{k \lambda r_{0}^2}{\beta }$, or $\mu >\frac{2 k \lambda
r_{0}^2}{\beta }$,  the first inequality is the condition
(\ref{eq:uneq1}), the second inequality must be discarded, because
it does not guarantee the absence of horizons in our solution.
Imposing the condition of absence of horizons result clear then,
that around the throat the null energy condition is defined positive
and tends asymptotically to zero when $r\rightarrow\infty$ as show
in the following Figure \ref{fig:rhop}.
\begin{figure}[!ht]
\begin{center}
\includegraphics[width=7cm]{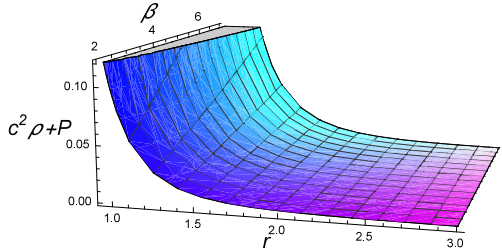}
\end{center}\vskip -5mm
\caption{The plot shows how the NEC is satisfied for different
values of $\beta$; considering $r_0=1$, $\lambda=.36$ and $\mu=1$. }
\label{fig:rhop}
\end{figure}
\section{Comments and Conclusions }\label{conclusiones}
In this paper we construct an exact solution for a wormhole from an
effective anisotropic fluid, using corrections of the braneworld to
gravitational 4D field equations. Our solution was built with the
help of an ansatz for the state equation of a perfect fluid 4D, it
should be noted that this solution meets the necessary conditions to
form a wormhole. In addition, it does not violate the NEC by the
perfect fluid for this particular solution. Also this solution
allows us to establish an astronomical bound for the radius of the
throat in terms of tension of the brane, for that reason worth
further explored the scope of ansatz used in this work in more
realistic cosmological scenarios.

\section{Acknowledgements}
We gratefully acknowledge support from CIC--UMSNH and RRML
acknowledge support from SNI--CONACyT.

\end{document}